\documentclass[twocolumn]{aa}

\usepackage[utf8]{inputenc}
\usepackage[english]{babel}
\usepackage{natbib}
\usepackage{graphicx}
\usepackage{amsmath}
\usepackage[varg]{txfonts}
\usepackage{setspace}
\usepackage{epstopdf}
\usepackage{tabularx}
\usepackage{ulem}
\usepackage{booktabs}
\usepackage{hyperref}

\institute{Hamburger Sternwarte, Universit\"at Hamburg, Gojenbergsweg
  112, 21029 Hamburg, Germany \label{inst1}}
\title{Star formation in evolving molecular clouds}
\titlerunning{Star formation in molecular clouds}
\author{M. V\"olschow \inst{\ref{inst1}}
\and R. Banerjee \inst{\ref{inst1}}
\and B. K\"ortgen \inst{\ref{inst1}}}
\authorrunning{V\"olschow et al.}

\offprints{{\tt marcel.voelschow@hs.uni-hamburg.de}}
\date{}

\DeclareMathOperator\erf{erf}

\def\D{\mathrm{d}}
\newcommand{\SFR}{\rm{SFR}}
\newcommand{\SFE}{\rm{SFE}}

\newcommand{\fsequa}{\, \, \, .}
\newcommand{\comequa}{\, \, \, ,}

\newcommand{\cm}{\rm{cm}}

\newcommand{\km}{\rm{km}}

\newcommand{\s}{\rm{s}}

\newcommand{\Msol}{{\rm M}_{\odot}}

\newcommand{\pc}{\rm{ pc}}

\newcommand{\K}{\rm{ K}}

\newcommand{\yr}{\rm{yr}}

\newcommand{\pcc}{{\rm ~cm}^{-3}}
\newcommand{\Myr}{\rm{Myr}}

\newcommand{\Ma}{{\cal M}}
\newcommand{\dd}{{\mbox{d}}}

\newcommand{\bef}{\begin{figure}[!t]}
\newcommand{\eef}{\end{figure}}

\abstract{
Molecular clouds are the principle stellar nurseries of our universe, keeping them in the focus of both observational and theoretical studies. From observations, some of the key properties of molecular clouds are well known but many questions regarding their evolution and star formation activity remain open. While numerical simulations feature a large number and complexity of involved physical processes, this plenty of effects may hide the fundamentals that determine the evolution of molecular clouds and enable the formation of stars. Purely analytical models, on the other hand, tend to suffer from rough approximations or a lack of completeness, limiting their predictive power.
In this paper, we present a model that incorporates central concepts of astrophysics as well as reliable results from recent simulations of molecular clouds and their evolutionary paths. Based on that, we construct a self-consistent semi-analytical framework that describes the formation, evolution and star formation activity of molecular clouds, including a number of feedback effects to account for the complex processes inside those objects.
The final equation system is solved numerically but at much lower computational expense than, e.g., hydrodynamical descriptions of comparable systems. The model presented in this paper agrees well with a broad range of observational results, showing that molecular cloud evolution can be understood as an interplay between accretion, global collapse, star formation and stellar feedback.
}  


\begin{document}

\maketitle 

\section{Introduction}
Molecular coulds, the birthplaces of stars in present day galaxies, are thought to be dynamical objects continously forming out of the diffuse interstellar medium \citep[see,e.g.][]{Blitz1980}. The most massive clouds may be referred to as \textit{giant molecular clouds}, but typically they contain masses between several hundred to several million solar masses and radii of a few to a few hundred parsecs. In order to form stars, one would expect molecular clouds to feature signatures of global collapse or at least some level of local instability.\\
At this point, the key question that arises is: how dynamic are molecular clouds? A widely used quantity to measure global stability is the virial parameter \citep[see,e.g.][]{Bertoldi1992}, taking into account the mass, size and velocity dispersion. \citet{Goldreich1974} argued that the linewidths observed in molecular clouds correspond to a state of global collapse. In contrast, authors such as \citet{Murray2011} found that typical GMCs have virial parameters of order unity, i.e. roughly maintain a quasi-static equilibrium, other authors such as \citet{Mckee2003} employed virialised cores as input for their turbulent core accretion model. \citet{Heyer2001} proposed that clouds up to $\sim 10^{4}~\Msol$ may be more stable, while larger clouds are subject to global collapse.\\
At the same time, works by \citet{Zuckerman1974} and \citet{Zuckerman1974b} showed that the simple picture of star formation via collapse and fragmentation of spherical clouds of cold dense gas leads to severe problems with the empirical basis. In fact, if all of our galaxy's dense molecular gas were in free-fall, the star formation rates (SFR) would be two orders of magnitude greater than observed \citep[see,e.g.][]{Krumholz2005}.\\
Today, observed SFRs of giant molecular clouds range between $1~\Msol/\Myr$ and $1000~\Msol/\Myr$ \citep{Lada2010} while canonical values for their star formation efficiency (SFE) lie in the range between a few percent for large clouds and a few ten percent for more compact clouds \citep[see,e.g.][]{Krumholz2005, Murray2011}. While small SFEs are generally attributed to large clouds and large SFEs to low-mass clouds \citep{Krumholz2005}, no such clear mapping exists for the SFR and different levels of star formation activity are found across a broad range of molecular cloud masses and sizes \citep{Murray2011}.\\ 
However, \citet{Palla2000} found that nearly all of the studied objects are the result of star formation processes that started at a low level, before a phase of steep acceleration set in. As a result, most of the stars form at the end of a molecular cloud's lifetime which is a few ten Myrs \citep[see,e.g.][]{Larson1981, Murray2011}, and we end up at age histograms that are dominated by young stars \citep[see,e.g.][]{Palla2000, Megeath2012}. The impact of turbulence, magnetic fields as well as the role of global collapse are still unclear and different authors favour different pictures \citep[see,e.g.][]{Klessen2000, Krumholz2005, Hennebelle2008, Zamora2012, Koertgen2015}.\\
Evidently, the field of star formation study is still dominated by ongoing fundamental debates and has to deal with a highly inconsistent empirical basis which supports a large number of different models. A modern ansatz that potentially solves the inconsistencies between theory and observations is the hybrid picture of a dynamical interplay between gas accretion, turbulence, global collapse, star formation and feedback \citep{Zamora2012}.\\
Early studies by \citet{Wannier1983} revealed envelopes of warm atomic gas surrounding molecular clouds. These warm gas envelopes are gravitationally bound to the molecular cloud and support continuous accretion, before feedback via winds, ionization and SNe finally destroys the surrounding envelope and ends accretion \citep[see,e.g.][]{Banerjee2009, Fukui2009}, naturally limiting the cloud's lifetime.\\ 
Large clouds may further suffer from so-called \textit{fragmentation-induced starvation}, limiting accretion onto individual stars and small stellar systems and the efficiency of star formation processes to the values we observe today \citep[see,e.g.][]{Peters2010, Girichidis2012}.
Colliding flows of galactic gas are a straight-forward explanation for the formation of gravoturbulent molecular clouds as they can initiate a phase transition from warm neutral gas to the cold neutral medium which has been studied intensively both numerically and theoretically \citep[see,e.g.][]{Hennebelle1999, Kojima2000, Vazquez2006}.
\citet{Vazquez1994} found that a lognormal distribution resembles the distribution of densities under isothermal gravoturbulent conditions which are indeed met in molecular clouds \citep[see,e.g.][]{Padoan2002,Krumholz2005, Hennebelle2008}. Observational studies by \citet{Schneider2013} or \citet{Schneider2014}, proved that such a \textit{lognormal} distribution of densities fits the observed density structure of giant molecular clouds very well, both in the solar neighborhood and towards the galactic center \citep[see][]{Rathborne2014}. More specifically, a work by \citet{Klessen2000} supported by recent observations \citep[see,e.g.][]{Schneider2014} and additional theoretical work \citep[see,e.g.][]{Federrath2012, Federrath2013, Girichidis2014} revealed the presence of a power-law tail at the high-mass end of the density distribution as an indicator of ongoing star formation processes.\\ 
Based on these recent findings, we propose a self-consistent dynamical model to describe the full evolution of molecular clouds on all scales via a set of analytical expressions. The model only depends on a small number of simple physical and astrophysical quantities frameworked by recent advances on how molecular clouds evolve, testing whether all these individual results can add up to one consistent model. In fact, our model is able to describe a large set of observations adequately and predicts a number of new effects that call for observational verification. The structure of our paper is as follows: first off, we start with the main ingredience and describe the way how we transform gas into stars. In the next section, we elaborate on the analytical framework how gas is accreted from surrounding HI streams and converted into a molecular cloud that evolves over time. Next, we explain our implementation of stellar feedback before we present and discuss the full model results.
\section{From gas to stars}
Any cloud models' main ingredience is how the dense parts of the cloud's density distribution are transformed into stars. Typically, the lognormal probability density function (PDF) as found in gravoturbulent molecular clouds is expressed in terms of the normalized logarithmic density coordinate $s:= \log ( \rho / \rho _{\rm 0} )$ with mean density of the cloud $\rho _{\rm 0}$. In this coordinate system, the canonical form of the volume-weighted PDF of density fluctuations is
\begin{equation}
P_{\rm V} (s) = \frac{1}{\sqrt{2\pi\,\sigma^2}}\, \exp \left( -\frac{(s-s_{\rm 0})^2}{2 \, \sigma ^2} \right)
\label{eq:PDF}
\end{equation}
with $s_{\rm 0} = -1/2 \, \sigma ^2$ \citep[see, e.g.][]{Vazquez2010} and a turbulence-dependend width $\sigma$ given by 
\begin{equation}
\sigma ^2 = \ln(1+b ^2 \, \Ma^2)
\end{equation}
where $\Ma$ is the thermal Mach number \citep[see e.g.,][]{Padoan2002, Federrath2010}. The parameter $b$ accounts for the type of turbulence \citep[see][]{Federrath2008}. \citet{Vazquez2010} set it to unity. Regarding the turbulent Mach number we follow \citet{Heiles2003} who showed that the coldest and densest parts of the ISM have typical mean Mach numbers of $3$ which we will use whenever explicit calculations are in order. Further, we impose a constant mean Mach number over the entire cloud evolution time, neglecting possible energy and momentum by supernovae, winds and outflows \citep[see,e.g.][]{Banerjee2007, Rogers2013, Bally2016, Padoan2016}. The PDF is normalized to
\begin{equation}
\int \limits _s ^{\infty} \dd s \, s \, P_{\rm V} (s) = 1 \fsequa
\end{equation}
\subsection{SFR and SFE}
\label{sec:sfr}
Formally, the instantaneous mass in stars $M_S(t)$ is related to the star formation rate $\SFR (t)$ via:
\begin{equation}
M_{\rm S} (t) = \int \limits _0 ^{t} \SFR(t') \, \D t' \fsequa
\end{equation}
Stars form out of collapsing fragments in the cold and dense parts of the cloud. \citet{Krumholz2005} found that the star formation rate is determined by the mass fraction of cold and dense gas over the free-fall time of the cloud which is $t_{\rm{ff}} = \sqrt{3 \pi / 32 G \mu m_{\rm H} n}$ in the spherical case or $t_{\rm{ff}} = \sqrt{R_{\rm C} / 2 h_{\rm C} G \mu m_{\rm H} n}$ with cloud radius and height $R_{\rm C}$ and $h_{\rm C}$ for cylindrical sheets \citep{Toala2012}. Using that, we can write the instantaneous star formation rate as
\begin{equation}
\label{eq:sfr_def}
\SFR(t) = \frac{M_{\rm thresh}}{t_{\rm ff}} = \frac{M_{\rm C} \, f}{t_{\rm ff}}
\end{equation}
where $M_{\rm thresh}$ is the star-forming mass-fraction and $\rm{f}$ denotes the fraction of the cloud's dense gas mass $M_{\rm C}$ that goes into stars.\\
Calculating $M_{\rm thresh}$ is the key aspect of this work. Different star formation criteria have been put forward in the literature to calculate the fraction of star-forming mass $\rm{f}$ \citep[see,e.g.][]{Padoan2002, Krumholz2005,Hennebelle2008}. We shall start with our new approach: a revised version of the \citep{Padoan1995} formalism. Our description closely follows \citet{Banerjee2014}.
\subsection{Star formation criteria}   
\subsubsection{The core-mass function formalism}
\label{sec:cmf}
\bef
\centering
\resizebox{\hsize}{!}{\includegraphics{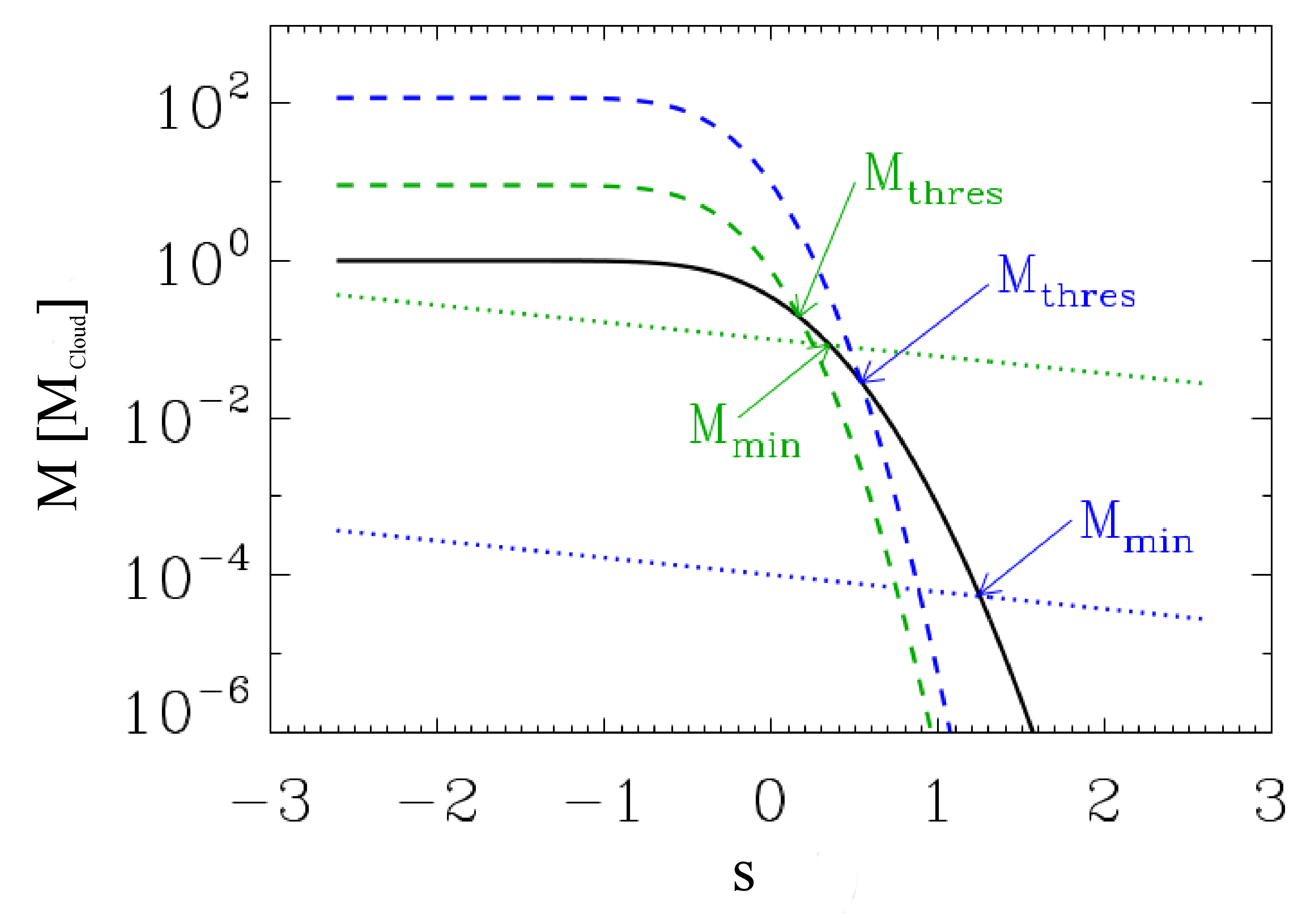}}
\caption{Illustration of the CMF formalism as described in sec.~\ref{sec:cmf}. The density of the lowest possible core can be interpreted as the intersection of the cumulated density PDF (black line) and the Jeans mass (dotted lines) which has a constant slope in this log-log plot. The dashed lines are the normalized Salpeter core-mass functions and their intersection with the cumulated density PDF equals the lower density threshold. $M_{\rm thresh}$ is the total mass that will be transformed into stars.}
\label{fig:cmf_formalism}
\eef
From observations, the distribution of stellar cores (core-mass function CMF) is known to follow a Salpeter power-law of the form
\begin{equation}
\rm{CMF} = \frac{\D N}{\D \log M} = C \, M^{-\alpha} 
\end{equation}  
with normalization constant $C$ and slope $\alpha$ \citep[see,e.g.][]{Salpeter1955, Lada2008, Rathborne2009}.\\
The smallest core that can be formed out of the cloud gas can be illustrated as a sphere of all the densest parts of the cloud cumulated into one single object, starting from some density threshold $s_{\rm min}$ and ranging to infinity. It collapses once its mass exceeds a Jeans mass $M_J (s_{\rm min})$. We can find this minimal core by solving
\begin{equation}
\frac{M_{\rm C}}{2} \left[1 - \erf{\left(-\frac{s_{\rm min}-\sigma^{2}/2}{\sqrt{2}\,\sigma}\right)} \right] = M_{\rm J} (s_{\rm min})
\end{equation}
where $M_{\rm C}$ is the cloud mass, i.e. integrating the dense parts of the PDF up to a density $s_{\rm min}$ where the total integrated mass exceeds a Jeans mass $M_J (s_{\rm min})$. Such a minimal core of mass $M_J (s_{\rm min}) = M_{\rm min}$ defines the high-mass end of the CMF. The CMF normalization parameter is determined via the total mass of the cloud:
\begin{equation}
\int \limits _{M_{\rm min}} ^{M_{C}} \D M M^{-\alpha} = C \frac{M_{\rm min}}{\alpha - 1} ~~~ \fsequa
\end{equation}
The normalized CMF can be used to calculate the number of cores within a mass range starting from an arbitrary mass threshold $M$ up to the cloud mass via
\begin{equation}
\begin{aligned}
N(M) 	&= C \, \int \limits _{M} ^{M_{\rm C}} \D M M^{-\alpha - 1} \\ 
	&= \frac{\alpha-1}{\alpha}\, \left(\frac{M_{\rm C}}{M_{\rm min}}\right)\,\left[\left(\frac{M}{M_{\rm min}}\right)^{-\alpha} - \left(\frac{M_{\rm C}}{M_{\rm min}}\right)^{-\alpha} \right] ~~~ \fsequa
\end{aligned}
\end{equation} 
Using the equation above, we can define a threshold mass by imposing that the mass fraction between $M_{\rm thresh}$ and $M_{\rm C}$ (see fig.~\ref{fig:cmf_formalism}) be at least one cloud which, in a hierarchical cloud picture is the largest cloud that contains all other cores \citep[see,e.g.][]{Vazquez2006}. This implies that we have to solve   
\begin{equation}
\label{eq:cmf_NM1}
N(M_{\rm thresh}) = 1
\end{equation}
for $M_{\rm thresh}$ (see fig.~\ref{fig:cmf_formalism}). After a few manipulations, we arrive at a total star-forming mass of
\begin{equation}
M_{\rm thresh} = M_{\rm min} \, \left( \frac{\alpha}{\alpha-1} \, \frac{M_{\rm min}}{M_C} + \left( \frac{M_{\rm min}}{M_{\rm C}} \right)^{\alpha}  \right)^{-1/\alpha}
\end{equation}
where a typical value for $\alpha$ is given by $1.35$ \citep[see,e.g.][]{Lada2008}. Using this mass, we can calculate the instantaneous SFR as given by eq.~\ref{eq:sfr_def} and define an instantaneous star formation efficiency $\rm{SFE} = M_{\rm thresh}/M_{\rm C}$ or
\begin{equation}
\label{eq:cmf_sfe}
\SFE =\left[ 
 \left(\frac{\alpha}{\alpha-1}\right)\, 
  \left(\frac{M_{\rm C}}{M_{\rm min}}\right)^{-1} 
+ \left(\frac{M_{\rm C}}{M_{\rm min}}\right)^{-\alpha}
\right]^{-1/\alpha}  \,
  \left(\frac{M_{\rm C}}{M_{\rm min}}\right)^{-1} ~~~ \fsequa
\end{equation}
Interestingly, the SFE only depends on the slope of the core-mass function and the cloud-core ratio $M_{\rm C}/M_{\rm min}$. In fig.~\ref{fig:cmf_formalism_sfe} we plotted the SFE as a function of the cloud-core ratio for two different CMF slopes. While massive clouds are expected to have SFEs in the range of a few percent, small clouds are much more efficient at forming stars.
\bef
\centering
\resizebox{\hsize}{!}{\includegraphics{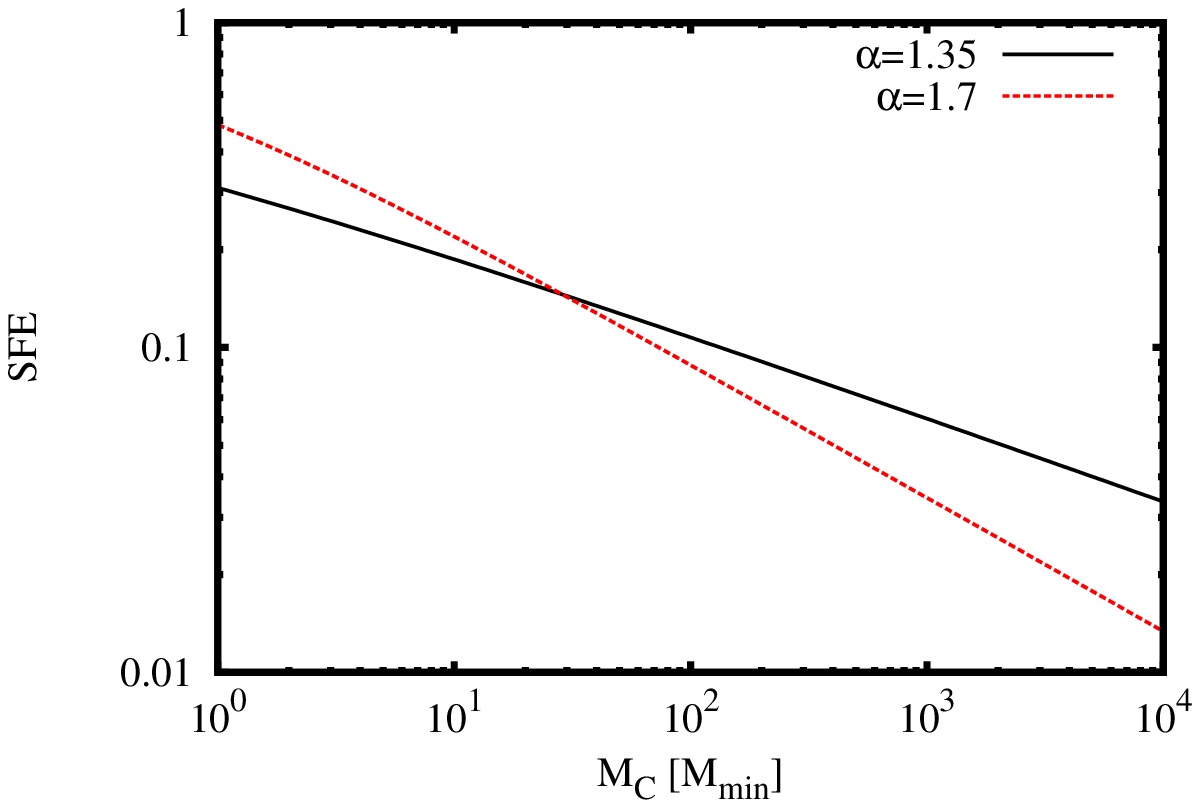}}
\caption{Star formation efficiency, as defined in sec.~\ref{sec:cmf} as a function of the cloud mass to minimal core mass ratio $M_{\rm C}/M_{\rm min}$. Smaller clouds have a typical $\SFE$ exceeding $10~\%$ while bigger clouds transform gas into stars less efficiently.}
\label{fig:cmf_formalism_sfe}
\eef
Compared to the \citet{Krumholz2005} criterion, our formalism does not depend on a set of efficiency factors determined via simulations. Instead, the only input parameters are the density PDF and the slope of the core mass function.
\subsubsection{Constant density threshold}
A number of authors argue that all mass beyond a specific constant density threshold is transformed into stars \citep[see,e.g.][]{Krumholz2005, Zamora2012}.\\
From the observer's side, the idea that star formation takes place in the densest parts of the cloud is a well-established fact and authors such as \citet{Lada2009} or \citet{Howard2014} give typical densities of star forming regions (clumps) in the order of $n_{\rm thresh} \simeq 10^{4}~\pcc$ and above. Explicitly, for a given star formation density threshold
\begin{equation}
s_{\rm thresh} = \ln \left( \frac{n_{\rm thresh}}{n_0} \right)
\end{equation}
the fraction of mass that goes into stars can be written as
\begin{equation}
f = \frac{1}{2} \left[1 - \erf{\left(-\frac{s_{\rm thresh}-\sigma^{2}/2}{\sqrt{2}\,\sigma}\right)} \right] \comequa
\end{equation}
assuming a log-normal PDF. While authors such as \citet{Zamora2012} fitted $s_{\rm thresh}$ to resemble numerical results, \citet{Krumholz2005} derived
\begin{equation}
s_{\rm thresh} = 1.07 \Ma _{\rm rms}^{2}
\end{equation} 
where $\Ma _{\rm rms}$ denotes the turbulent Mach number of the cloud.
Authors such as \cite{Zamora2012} found ranges between $10^{5}~\pcc$ and $10^{7}~\pcc$ as best-fit solutions\footnote{Note that the authors employ a different definition of the SFR that incorporates the local core free-fall time rather than the global cloud free-fall time.} while observations tend to threshold values in the range of $10^{4}~\pcc$ to $10^{5}~\pcc$ \citep[see,e.g.][]{Pudritz2013, Howard2014}. Studies of molecular clouds in the galactic center show increased star formation threshold densities which does not support the idea of a single universal density threshold \citep[see, e.g.][]{Kruijssen2014}. Further, the constant-density criterion shows purely positive scaling for increasing cloud turbulence (cf. fig.~\ref{fig:mach_vary}), contrary to findings by authors such as \citet{Krumholz2005}. On top of that, the total mass beyond such extremly high density thresholds may even be smaller than a single thermal Jeans mass. The differences between the three criteria presented here are subject of the next subsection.
\subsection{Comparing different star formation criteria} 
\label{sec:comparison}
\subsubsection{Threshold and core densities}
In what follows, we calculate typical star formation threshold densities and minimal core densities in the CMF formalism for a set of different clouds. Fig.~\ref{fig:densities_1e2} shows the SF threshold densities and minimal core densities for clouds of masses between $10^{2}~\Msol$ and $10^{6}~\Msol$, all with a mean density of $100~\pcc$ and a molecular weight $\mu = 2.35$. We assumed $\Ma = 3$ for all clouds.
\bef
\centering
\resizebox{\hsize}{!}{\includegraphics{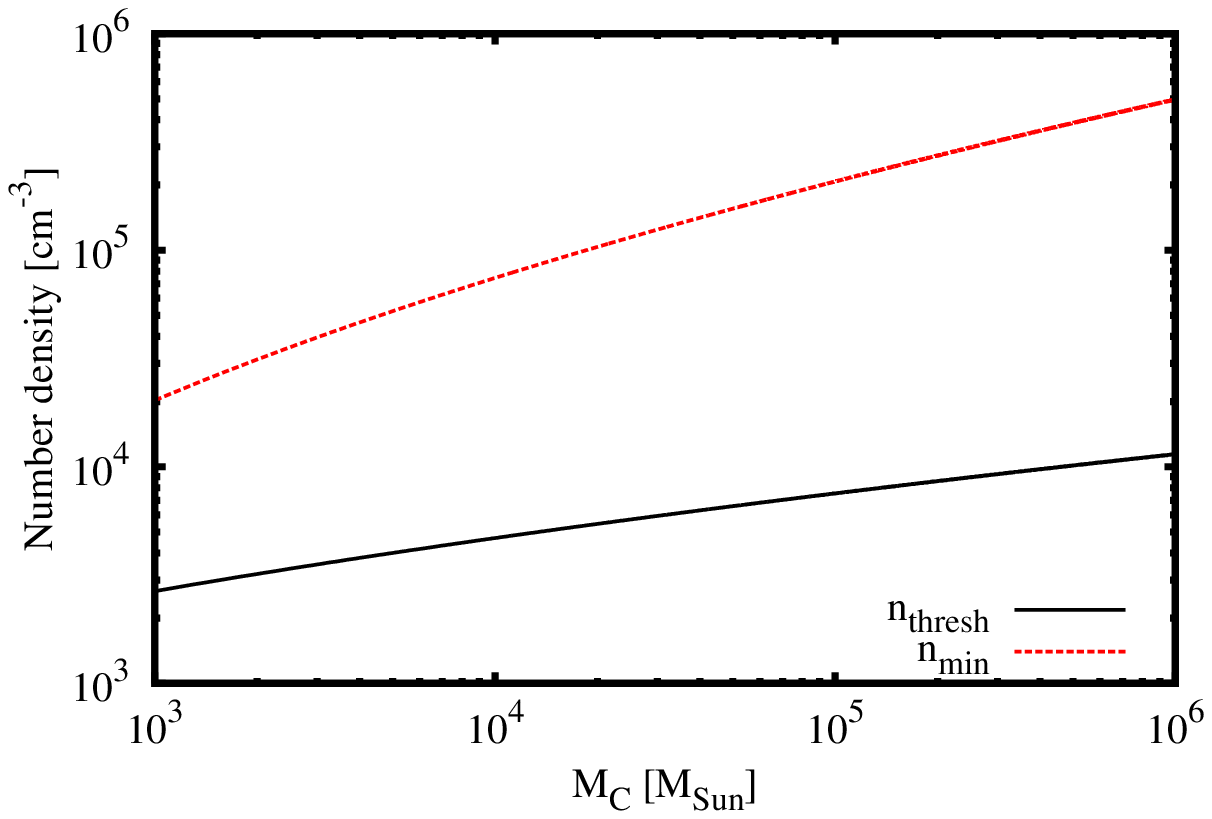}}
\caption{Minimal core density and star formation threshold density for a cloud with a mean density of $100~\pcc$ and varying mass. For a given cloud, all mass between those two thresholds will be converted into stars.}
\label{fig:densities_1e2}
\eef
For the typical mass and density range presented in fig.~\ref{fig:densities_1e2}, the minimal core densities lie between $10^{5}~\pcc$ and $10^{8}~\pcc$. The SF threshold densities vary between $10^{3}~\pcc$ and $10^{6}~\pcc$ for extremely dense and massive clouds bracketing the typical star formation threshold densities found in the literature (see sec.~\ref{sec:cmf}).\\ 
\subsubsection{Star formation rates and efficiencies}
Fig.~\ref{fig:sfr_1e2} shows the star formation rate for clouds of varying mass and identical mean density of $100~\pcc$, comparing the CMF formalism and different constant threshold densitiy criteria. We assumed $\Ma = 3$ for all clouds.
\bef
\centering
\resizebox{\hsize}{!}{\includegraphics{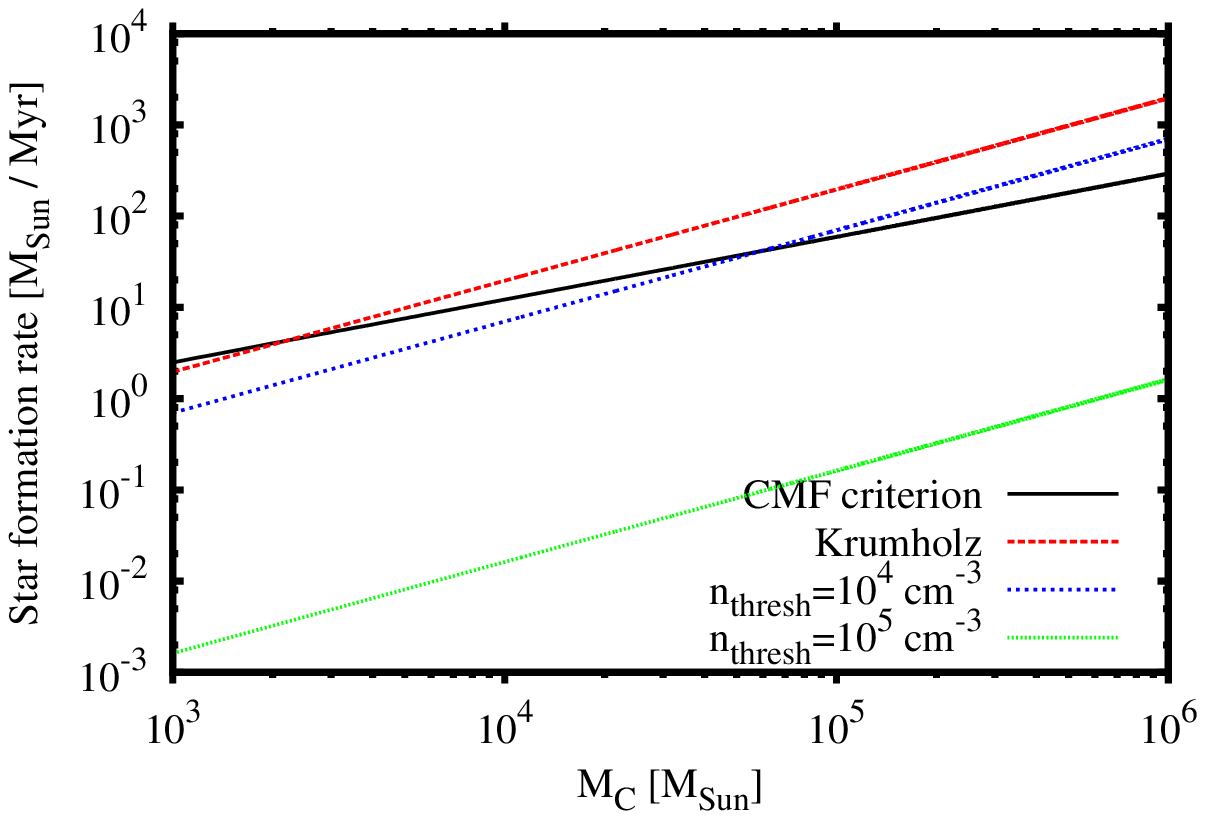}}
\caption{Star formation rates for a cloud with a mean density of $100~\pcc$ and varying mass, calculated with different star formation criteria: The CMF formalism, the \citet{Krumholz2005} formalism and two constant-threshold criteria. In the high-threshold case ($10^{5}~\pcc$), virtually no stars are formed even for massive clouds while all other criteria yield comparable SFRs and show similar scaling.}
\label{fig:sfr_1e2}
\eef
In fig.~\ref{fig:sfr_1e2}, the star formation rates given by the different star formation criteria spread over eight orders of magnitude. The CMF criterion, the \citet{Krumholz2005} formalism and the low-density threshold predict similar star formation rates and show comparable scaling. On the other hand, for the $10^{5}$ threshold criterion we expect almost no star formation at all with star formation rates far below $1~\Msol / \Myr$, conflicting with observational results for such low-mass molecular clouds which are indeed actively forming stars \citep[see,e.g.][]{Murray2011}.
\subsubsection{The impact of turbulence}
\label{sec:turbulence}
In fig.~\ref{fig:mach_vary}, we selected a cloud with intermediate properties, i.e. a density of $10^{3}~\pcc$ and a mass of $10^{5}~\Msol$, and varied the mean Mach number of the cloud to work out the impact of turbulence on the star formation rate.
\bef
\centering
\resizebox{\hsize}{!}{\includegraphics{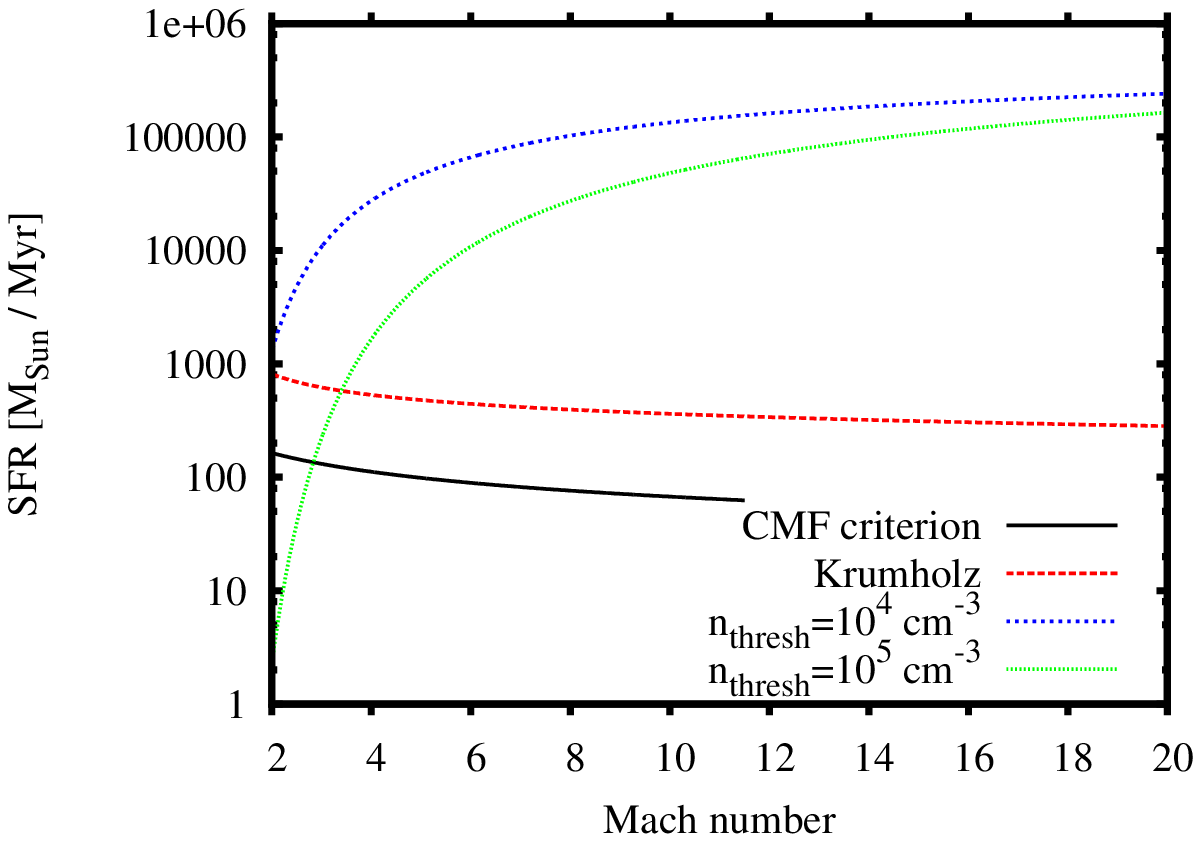}}
\caption{Turbulence dependence of the star formation rate for an intermediate cloud with mean density $10^{3}~\pcc$ and mass $10^{5}~\Msol$. Constant threshold density models are very sensitive to changes in the cloud's global Mach number with a clearly positive trend for increasing Mach numbers. On the other hand, the star formation rate as calculated via the CMF formalism scales just weakly with the Mach number and gives results similar to the \citet{Krumholz2005} criterion. Both criteria show opposite scaling: additional turbulence decreases star formation rates. As explained in sec.~\ref{sec:turbulence}, in the CMF formalism star formation is terminated beyond a certain Mach number threshold.}
\label{fig:mach_vary}
\eef
As expected, the constant-threshold criteria predict higher star formation rates for increasing Mach numbers. This result represents the fact that higher Mach numbers lead to a broader PDF with more mass at lower and higher densities. The effect is dramatic: for the high-threshold criterion, the star formation rate increases by three orders of magnitude if we increase the Mach number from $2$ to $6$, i.e. by half a magnitude. The lower thresholds react less sensitive to Mach number variations but still, constant threshold density star formation criteria are sensitive to changes in the cloud's mean Mach number with a purely positive feedback on the star formation rate.\\
In contrast, the CMF criterion shows almost no variation of the star formation rate for increasing Mach numbers, with just a weakly negative trend resulting from the broadening of the PDF for increasing Mach numbers.
\bef
\centering
\resizebox{\hsize}{!}{\includegraphics{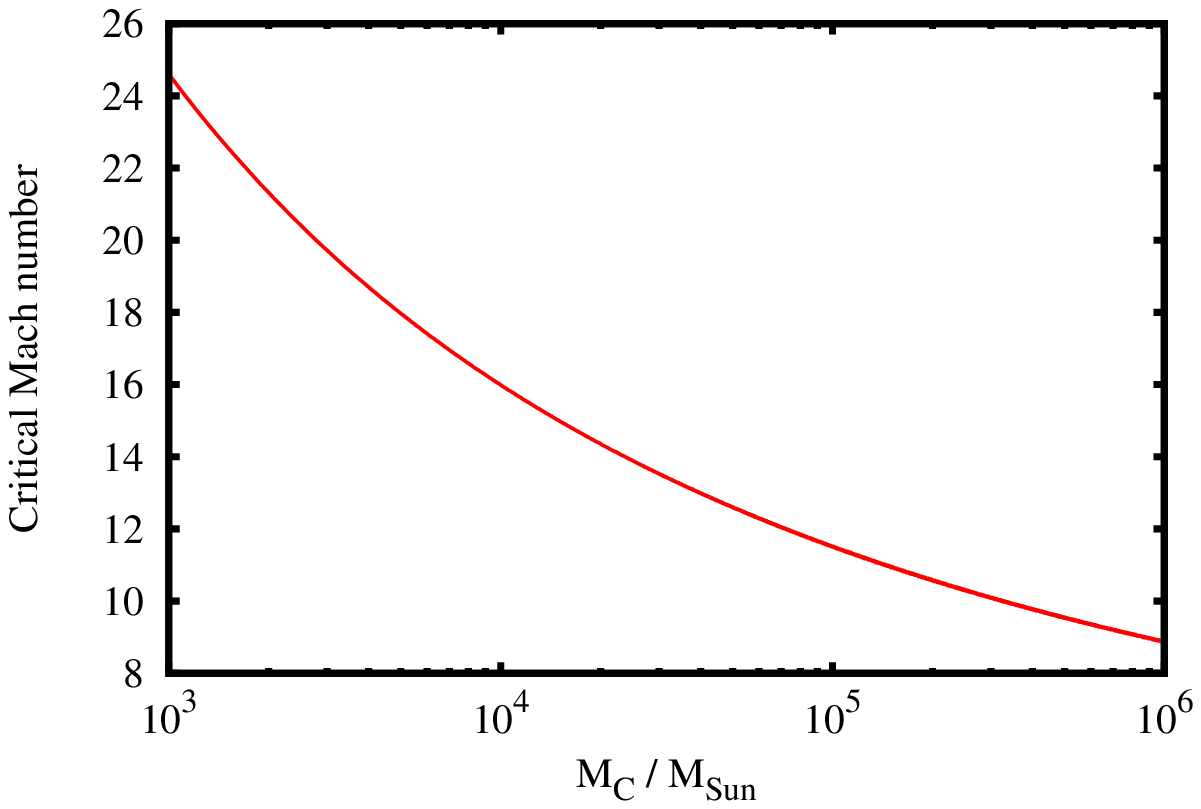}}
\caption{Critical Mach number as a function of the cloud's mass for a cloud with mean density $10^{3}~\pcc$. Smaller clouds continue their star formation activity even for very high Mach numbers, while larger clouds with more than a few $10^{5}~\Msol$ terminate below Mach $10$.}
\label{fig:cmf_mach_cutoff}
\eef
However, beyond a Mach number of $11$, eq.~\ref{eq:cmf_NM1} has no solution and the $\SFR$ drops to zero. Hence, the CMF criterion predicts that star formation is terminated beyond a critical Mach number (see fig.~\ref{fig:cmf_mach_cutoff}). Below, the CMF criterion and the \citet{Krumholz2005} threshold show similar scaling and differ only by half a magnitude. 
\section{Cloud evolution}
\label{sec:cloud_evo}
In what follows, we describe a full dynamical model framework inspired by \citet{Zamora2012} covering accretion, contraction, star formation and feedback in which we implemented the CMF star formation criterion.
\subsection{Model framework}
Let's assume that the gaseous part of the cloud ($M_{\rm C}$) follows a mass budget equation of the form 
\begin{equation}
M_{\rm C} (t) = M_{\rm A} (t) - M_{\rm S} (t) - M_{\rm F} (t) \fsequa
\end{equation}
The cloud accretes gas $M_{\rm A}$ which increases its gas content, while formation of stars $M_{\rm S}$ and feedback effects ($M_{\rm F}$) may decrease the gas content again.\\
The most fundamental process in order to form a molecular cloud is accretion from the surrounding medium via gas streams or HI envelopes \citep[see, e.g.,][]{Vazquez2006, Fukui2009}. 
\subsection{Accretion}
\label{sec:accretion}
Assuming that the cloud's geometrical radius $R_{\rm C}$ defines its cross-section for the accretion process and the cloud is fed by two streams, the mass infall rate is given by  
\begin{equation}
\dot{M}_{\rm inf}(t) = 2 \, \rho _{\rm inf} \, v_{\rm inf} \, ( \pi R_{\rm C}^{2} ) \comequa
\end{equation}
with infall gas density $\rho _{\rm inf}$ and gas velocity $v_{\rm inf}$ \citep[see][]{Zamora2012}, where the mass density $\rho$ is linked to the number density $n$ via $\rho _{\rm inf} = n_{\rm inf} \, \mu _{\rm H}$ with mean molecular weight $\mu$. As our model focusses on the evolution of the cold medium and does not cover the details of the transformation from atomic to molecular gas, we require the cloud to be shielded against the ambient UV flux. Given the initial mean density of the cloud of $100~\cm^{-3}$ and the size of the smallest clouds we consider (a few pc), typical shielding column densities are well exceeded \citep[see,e.g.][]{Franco1986}. The instantaneous accreted mass now reads
\begin{equation}
M_{\rm A} (t) = \int \limits _0 ^{t} \dot{M}_{\rm inf} (t') \, \D t' \comequa
\label{eq:accreted_mass}
\end{equation}
where we assume that the warm gas from the streams is directly converted into molecular gas $M_{\rm C}$. Out of this molecular cloud gas, stars form as given by eq.~\ref{eq:sfr_def} where we calculate the fraction of mass that goes into stars using our CMF criterion (cf. sec.~\ref{sec:cmf}). In addition, following \citet{Zamora2012} we define an instantaneous star formation efficiency
\begin{equation}
\rm{SFE} = \frac{M_{\rm S}}{M_{\rm Acc}}
\end{equation}
as the mass in stars over the total accreted stream gas mass, i.e. mass in dense gas, diffuse gas and stars.\\
Throughout the evolution, we keep the initial cloud radius as accretion radius accounting for gravitational focussing. Note that this model does not depend on a colliding flow scenario but for the moment we keep the comprehensive picture of two feeding streams. Further, we did not account for accretion flow feedback which can terminate the accretion of new material for weak accretion flows \citep[see,e.g.][]{Peters2010}.
\subsection{Global evolution}
For the colliding streams scenario, \citet{Vazquez2006} found that the cloud's global evolution can be divided up into two stages:
\begin{itemize}
\item Accretion stage: the cloud only grows in height at a constant radius and constant density until it becomes Jeans-unstable.
\item Contraction stage: the cloud's density and height remain approximately constant, but radial contraction sets in. As a result, the cloud's radius decreases and its mean density increases.
\end{itemize}
During the two evolutionary stages, different boundary conditions for the cloud's evolution hold.
\subsubsection{Accretion stage}
Initially, the cloud only grows in height while maintaining a constant mean density of $n_{\rm 0} = 10^{2}\pcc$ and constant cloud radius $R_{\rm C} = R_{\rm inf}$, i.e. equalling the inflow radius. Under such boundary conditions, the identity
\begin{equation}
M_{\rm C} (t) = \rho _{\rm inf}  \, \pi R_{\rm C}^{2} \, h (t)
\end{equation}   
applies which can be solved for the cloud's height $h(t)$. The mass accretion stage ends once the cloud exceeds its thermal Jeans mass. For a finite and isothermal sheet of gas, \citet{Larson1985} gives a Jeans mass of
\begin{equation}
M_{\rm J} = 4.67 \frac{\pi R_{\rm C}^{2} c_{\rm s}^{4}}{G^{2} M_{\rm C}} \comequa
\end{equation}
where $c_{\rm s}$ denotes the mean speed of sound of the cloud which is given by $c_{\rm s} = \sqrt{k_{\rm B} \, T_{\rm C} / \mu \, m_{\rm H}}$ with Boltzmann constant $k_{\rm B}$, mean cloud temperature $T_{\rm C}$, mean molecular weight $\mu$ of the cloud and hydrogen mass $m_{\rm H}$. Following \citet{Zamora2012} with a given initial mean density of $100~\pcc$, the cooling function provided by \citet{Koyama2002} gives a mean cloud temperature of $T_{\rm C} = 42~\K$, constant throughout the accretion phase. Once the cloud exceeds its Jeans limit, the transition to the contraction stage occurs.
\subsubsection{Contraction stage}
During the contraction stage, the cloud maintains its height and starts to contract radially. Following \citet{Vazquez2010} and \citet{Zamora2012}, we use a coordinate system with origin in the center of the cloud and x-direction defined by one of the inflows. In this reference frame, we calculate the self-gravity of the cloud by integrating over mass elements $\bar{\rho} \D x  \D y \D z$ summing up to a gravitational acceleration of
\begin{equation}
a(t) = G \bar{\rho} \int \limits _{-R_{\rm C}} ^{+R_{\rm C}} \D z \int \limits ^{\sqrt{R_{\rm C}^{2}-z^{2}}} _{0} \D y \int \limits _{-h/2} ^{h/2} \frac{R_{\rm C} - z}{[(R_{\rm C}-z)^{2}+y^{2}+x^{2}]^{3/2}} \D x \fsequa
\end{equation}   
The first and second integral can be solved analytically. Finally, we arrive at
\begin{equation}
a(t) = 2 G \bar{\rho} \int \limits _{-R_{\rm C}} ^{+R_{\rm C}} \D z \arctan \left( \frac{h \sqrt{R_{\rm C}^{2}-z^{2}}}{(R_{\rm C}-z)\sqrt{h^{2}+8 R_{\rm C} (R_{\rm C}-z)})} \right) \fsequa
\end{equation}
This integral is solved numerically using the Simpson rule and the new cloud radius is calculated as described in \citet{Zamora2012}.\\
For a real cloud, collapse is usually not in pure free-fall as thermal, magnetic and turbulent pressure counteract gravity. In order to account for that, \citet{Zamora2012} introduced a correction factor $f_L$ based on Larson's work on molecular cloud collapse who found that typical collapse times are slower by a factor of $1.58$.\\
Further, we assume that the streams are gravitationally focussed onto the cloud, keeping the accretion radius and rate constant throughout the simulation.
\subsection{Mass function evolution}
In our model, we assume that every fraction of stars $\Delta M$ formed at a given timestep $\Delta t$ follows a Kroupa IMF. Typical lifetimes can be calculated using
\begin{equation}
\tau \simeq 10^{10}~\yr \left( \frac{M}{\Msol} \right)^{-\alpha}
\end{equation} 
with $\alpha \approx 3-4$, where we adopt $3.5$ \citep[see][]{Adams1997}. This allows for explicit calculation of the shape of the stellar mass function which is crucial for the calculation of the feedback effects as carried out in the next section.
\section{Stellar feedback} 
\subsection{Stellar Wind}
In the recent literature, two contradicting positions are represented regarding the role of stellar wind and its impact of star formation: no impact versus large impact. Authors such as \citet{Ngoumou2014} studied the impact of a massive O7.5 star on stellar cores finding that stellar winds have just modest direct effects on their formation and distribution. Instead, low and intermediate density gas shows the most significant modulations but as it does not contribute to the star formation processes, \citet{Ngoumou2014} conclude that the global impact of stellar winds on star formation is modest.\\
\citet{Rogers2013} studied the feedback effects of three massive stars embedded into a $3,240~\Msol$ molecular cloud. Consistently, they found that coupling between the stellar wind and the molecular material is weak. By the end of the lifetime of the massive stars, just a small fraction of the molecular cloud is pushed out of the simulation box, following paths of least resistance through the porous gas. Overdense regions remained largely unaffected.\\
\citet{Dale2014} studied the combined effects of ionization and stellar winds finding that overall star formation efficiencies are reduced just by $10-20~\%$. According to them, the main effect of stellar winds is creation of cavities with radii of several parsecs which modulate the efficiency of consecutive feedback processes. Low-mass clouds are more affected.\\
In contrast to the work mentioned previously, newer results by \citet{Fierlinger2015} re-emphasize the importance of stellar winds. In extensive 1D studies of the energy retained in the surrounding molecular cloud, the authors found that wind-blown cavities reduce radiative losses of subsequent supernovae and may increase the retained energy and thus the feedback efficiency. In general, the energy injected by stellar winds and by supernovae are of comparable order of magnitude.\\ 
As a conclusion, stellar winds and their interaction with molecular material are still not fully understood and all studies presented here have limitations in terms of neglected processes, limited resolution or size of the cloud and cluster. For the moment, we follow authors such as \citet{Rogers2013}, \citet{Dale2014} and \citet{Ngoumou2014} and neglect the impact of stellar winds. 
\subsection{Ionization}
\label{sec:feedback_ion}
%
%
An approximative treatment of ionization was proposed by \citet{Zamora2012}: the authors calculated the mass of a representative typical massive star, i.e. the mean mass of the Kroupa IMF in the range $[8,60]~\Msol$ which is $\langle M_{OB} \rangle = 17~\Msol$ with its corresponding Lyman flux of $\bar{S}_\ast = 2 \cdot 10^{48}~\s^{-1}$, and applied this flux to the total number of massive stars. In our model, we assume that the gravoturbulent conditions and the constant mass inflow allow for efficient recombination. Hence, the total ionized mass is given by the gas inside all massive stars' Stroemgren spheres, i.e. 
\begin{equation}
R_{\rm{HII},0} = \left( \frac{3 \, \bar{S}_\ast}{4 \pi \, \alpha^{\ast}_{\rm H} \bar{n}^2} \right)^{1/3} \fsequa
\end{equation}
Here, $\bar{S}_{\ast}$ denotes the star's UV flux, $\alpha^{\ast}_{\rm H}$ is the hydrogen recombination coefficient and $n$ the mean number density of the cloud. Following \citet{Zamora2012} and assuming a gas temperature inside the HII regions of $10^{4}~\K$, we have a recombination coefficient of $\alpha^{\ast}_{\rm H} = 2.6 \cdot 10^{-13} \pcc \s$ and arrive at typical Stroemgren radii in the range of a parsec.
Over time, the heated HII bubbles expand into the molecular material which can be described via
\begin{equation}
R_{\rm{HII}} = R_{\rm{HII},0} \left( 1 + \frac{7 \, c_{\rm{HII}} \, t}{4 \,  R_{\rm{HII},0}} \right) ^{4/7}
\end{equation}
\citep[see][]{Spitzer1978}. $c_{\rm{HII}}$ is calculated assuming a gas temperature of $10^{4}~\rm{K}$ and $\mu = 0.5$. For the age of the HII region $t$ we calculate the mean age of all massive stars that formed up to the given time. In order to account for the expected amount of overlap between the HII regions, we assume an equal 2D distribution of the (massive) stars in the y-z plane and apply an overlap correction factor of the form
\begin{equation}
M_{\rm{ion}} = M_{\rm C} \, \left[ 1 - \exp \left( -\beta \, N_{\rm{HII}} \right) \right]
\end{equation}
where $\beta$ is a function of $\xi = R_{\rm{HII}} / R_{\rm C}$. In extensive simulations of clouds with a varying number of equally-sized HII regions, we calculated the mean fractional area covered by HII regions. We find that $\beta = 0.830865 \, \xi^{2.01978}$ fits the simulated data well.
\subsection{Supernovae}
Massive stars beyond a certain mass threshold (typically $8~\Msol$) end their life with a core-collapse supernova inserting typical energies of $10^{51}~\rm{erg}$ into their environment within short times \citep[see,e.g.][]{Koertgen2016}. The expanding shockwave destroys molecular material, potentially affecting star formation processes as well as the global gas dynamics. Authors such as \citet{Rogers2013} studied the impact of supernovae on a molecular environment finding that the the expanding supernova shockwave indeed destroys the molecular material it bypasses within a radius of several parsecs but does not fully evacuate the affected region because of weak coupling. Rather, the heated material may cool quickly and re-form molecular gas. If gas dynamics allow for efficient cooling and transport. As a result, more than $99~\%$ of the supernova energy may escape the cloud \citep{Rogers2013}.\\
Based on these findings, we do not expect supernovae to influence the global dynamics of our cloud, i.e. the global mean Mach number or the distribution of densities. Thus, we limit the effect of supernova explosions to destruction of molecular gas within a radius of $R_{\rm SN} = 5~\pc$ consistent with typical cooling timescales of $10^{3}~\yr$ and shockwave velocities $\sim 10^{4}~\km \s ^{-1}$ \citep[see][]{Rogers2013} and results by \citet{Koertgen2016} and \citet{Wareing2017}. The SN frequency is calculated self-consistently using the number of massive stars that left the main sequence. Again, a statistical overlap correction is applied (see sec.~\ref{sec:feedback_ion}).
\section{Full model parameter study}
In the previous sections, we derived all the building blocks necessary to set up our full model in which we incorporated the CMF star formation criterion as described in sec.~\ref{sec:cmf}. In what follows, we present an overview of our simulations for clouds with varying initial radii. The parameters we use are summarized in tab.~\ref{tab:parameters}. Every cloud's evolution was tracked until it either dropped below a radius of $1~\pc$, exceeded a mean density of $10^{6}~\cm^{-3}$ or lost more than $99~\%$ of its mass by feedback processes.
\begin{table*}
\caption{Summary of parameters used throughout all calculations.}
\begin{center}
\begin{tabular}{lccc}
\toprule
Name 	& Symbol	& Value 		& Unit 	\\
\midrule
Salpeter core mass function slope	(linear)	& $\alpha$ & 2.35 & - \\
Hydrogen recombination coefficient		& 	$\alpha _{\rm H} ^{\ast}$  & $2.6~10^{-19}$	&	$\s^{-1}$ 	\\
Larson parameter			& 	$f_L$ 				&	$1.7$	& - 	\\
Kroupa IMF lower limit			& 	$M_{\rm{Kroupa,min}}$ 				&	$0.01$	& $\Msol$ 	\\
Kroupa IMF upper limit			& 	$M_{\rm{Kroupa,max}}$ 				&	$60$	& $\Msol$ 	\\
Cloud mean molecular weight			& 	$\mu _{C}$ 				&	$2.35$	& - 	\\
Inflow mean molecular weight			& 	$\mu _{\rm{inf}}$ 				&	$1.27$	& - 	\\
Cloud mean Mach number			& 	$M _{rms}$ 				&	$3$	& - 	\\
Inflow number density (initial cloud density)	& 	$n_{\rm{CNM}} = n_{\rm 0}$  &	$100$	& $\pcc$ 	\\
Supernova ionization radius	& 	$R_{\rm{SN}}$  &	$5$	& $\pc$ 	\\
Inflow velocity	& 	$v_{\rm{inf}}$  &	$10$	& $\km / \s$ 	\\
Cloud temperature	& 	$T_{\rm C}$  &	$42$	& $\K$ 	\\
\bottomrule
\end{tabular}
\label{tab:parameters}
\end{center}
\end{table*}
\subsection{Temporal evolution}
Before we show and discuss the results of our full parameter study, we shall start with the temporal evolution of one generic cloud to illustrate a common evolutionary path. As a working example, we chose $R_{\rm 0} = 64~\pc$ (see fig.~\ref{fig:temp_evo}). The cloud continously accretes mass from the streams, leading to an increasing gas content. As a result, stars form at initially very low rates $< 10~\Msol/\Myr$ but with a clearly accelerating trend. Consequently, the stellar content increases and so do the feedback effects. Initially, feedback via SNe and ionization are of comparable effect but the second half of the cloud's lifetime is clearly dominated by SN feedback. Finally, SN feedback disrupts the cloud and star formation is terminated.
\bef
\centering
\resizebox{\hsize}{!}{\includegraphics{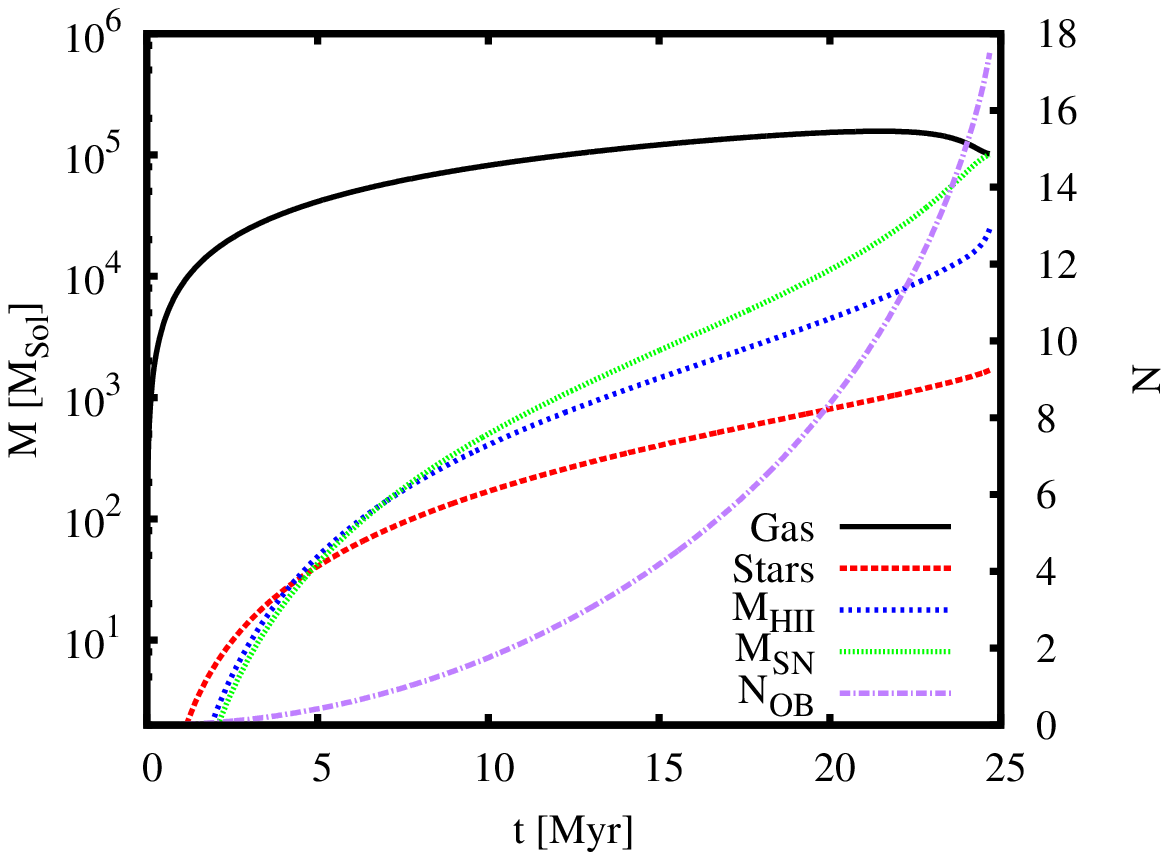}}
\caption{Temporal evolution of a typical cloud. Here, we selected $R_{\rm 0} = 64~\pc$ as an example setup and plotted the evolution of the gas content, stellar content as well as the amount of gas destroyed via ionization and supernova feedback.}
\label{fig:temp_evo}
\eef
Compared to the \citet{Zamora2012} model on which we based our work, feedback builds up much slower and is less dominant, leading to higher cloud masses (their fig. 2). As a consequence of both our lower overall feedback efficiency and our higher fiducial flow velocity, the lifetime of our cloud is reduced as is the final stellar content.
\subsection{Peak gas and star masses}
\bef
\centering
\resizebox{\hsize}{!}{\includegraphics{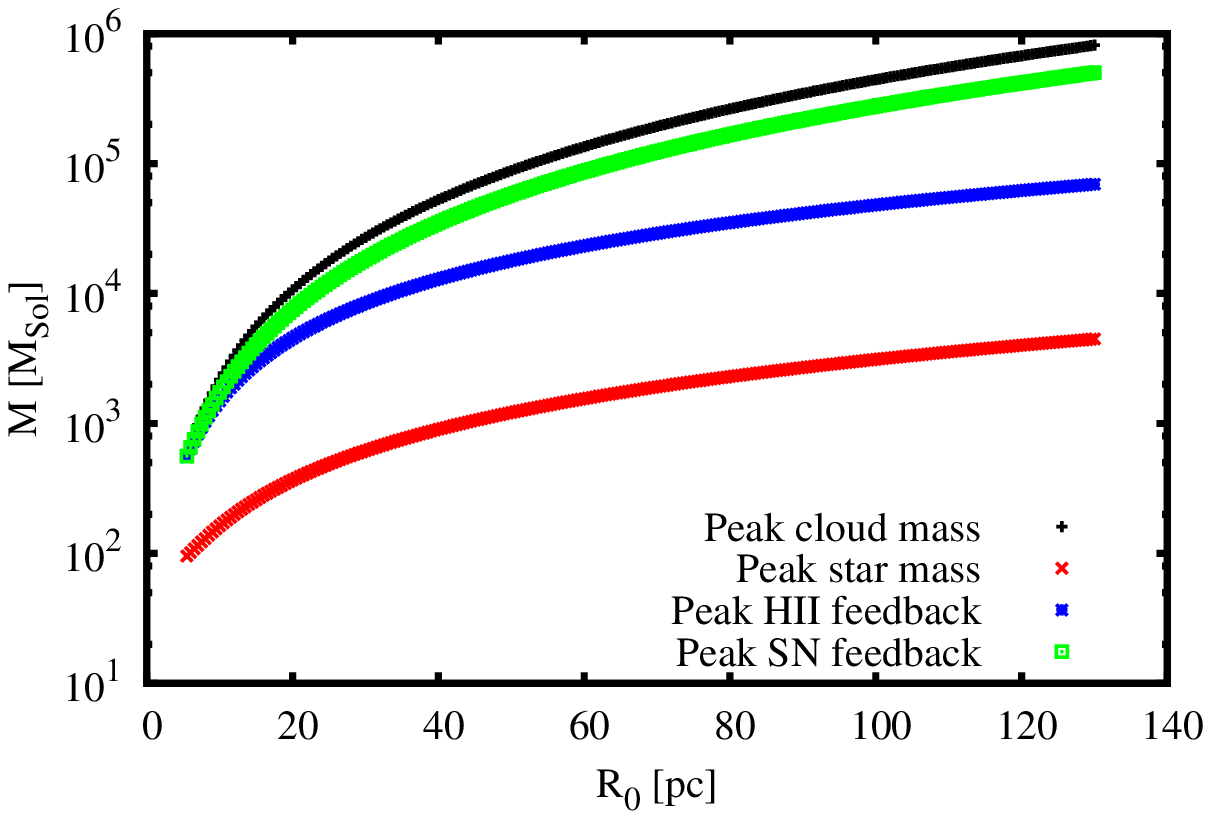}}
\caption{Peak gas mass, star mass and mass destroyed via feedback processes for clouds with varying initial radii. Throughout the parameter range, larger initial cloud and accretion stream radii lead to larger peak gas masses, an increased stellar content and a higher fraction of mass destroyed by feedback processes. The peak gas mass and star mass follow simple power laws.}
\label{fig:parstud_peak_mass}
\eef
In fig.~\ref{fig:parstud_peak_mass} we present four peak masses as functions of the initial cloud radius: gas and stellar content as well as the peak ionized mass and mass destroyed by supernovae. All four quantities scale clearly positively with initial cloud radius. Supernova feedback is the dominating feedback process, only for small clouds and during the first half of the cloud lifetime ionization is of comparable order of magnitude. Specifically, we find that the peak gas mass can be described with a generic power law of the form 
\begin{equation}
\frac{ \hat{\rm{\rm M}}_{\rm C}}{\Msol} \simeq 10 \left( \frac{R_{\rm 0}}{\pc} \right)^{2.3}
\end{equation}
while the peak star mass follows 
\begin{equation}
\frac{ \hat{\rm{M}}_{S}}{\Msol} \simeq 6.0 \left( \frac{R_{\rm 0}}{\pc} \right)^{1.4}
\end{equation}
with standard errors for both parameters well below one percent. Throughout the entire parameter space, all clouds take their peak mass and peak star mass at the end of the cloud lifetime, just before a dramatic drop in gas content and cloud disruption by feedback. Thus, the peak star mass and gas mass can be combined to a relation which reads
\begin{equation}
\frac{ \hat{\rm{M}}_{S}}{\Msol} \simeq 1.6 \left( \frac{\hat{\rm{M}}_{C}}{\Msol} \right)^{0.58} \fsequa
\end{equation}

\subsection{Star formation rates and efficiencies}
\label{sec:parstud_sfre}
\bef
\centering
\resizebox{\hsize}{!}{\includegraphics{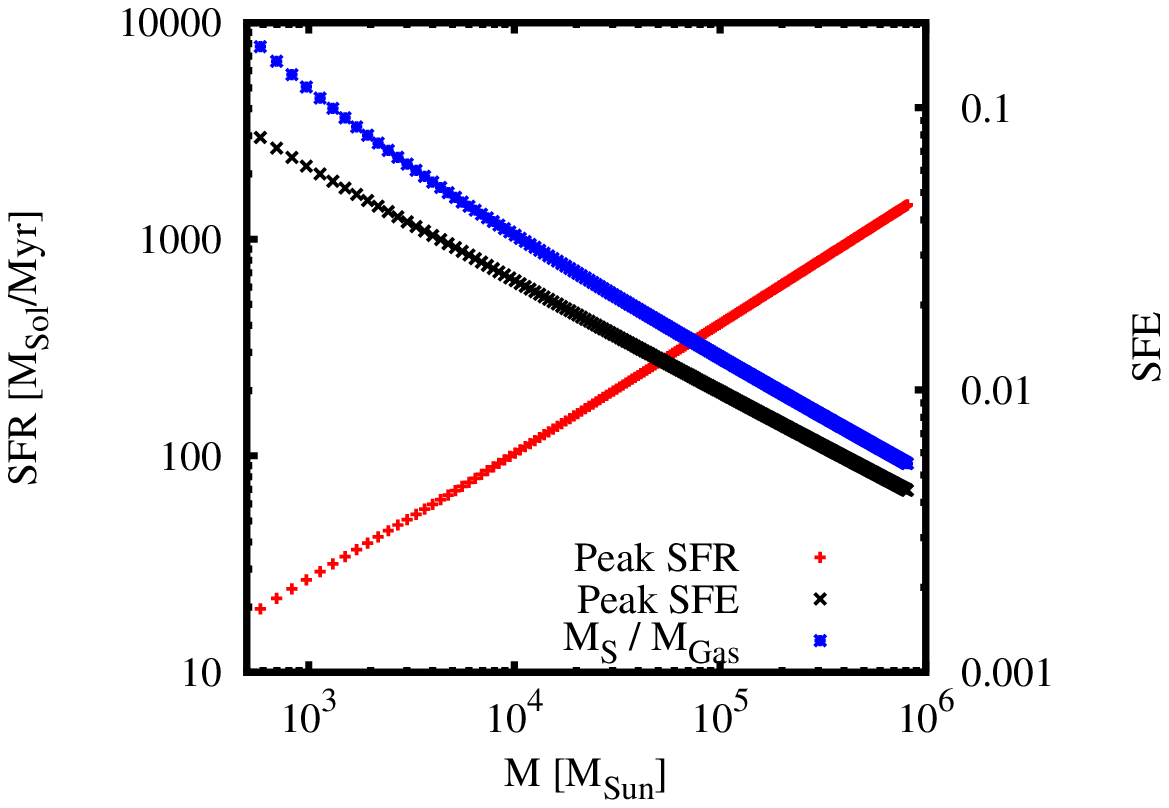}}
\caption{Peak star formation rate, peak star formation efficiency and peak star mass over peak gas mass plotted as functions of the peak cloud mass. Both the peak SFR and SFE show a simple power-law scaling.}
\label{fig:parstud_sfre}
\eef
Fig.~\ref{fig:parstud_sfre} shows the peak SFR and SFE of clouds with varying peak cloud mass. Both the SFE and the SFR have a power-law scaling. We find that the power laws
\begin{equation}
\frac{\hat{\SFR}}{\Msol/\Myr} \simeq 0.40 \left( \frac{\hat{\rm{M}}_{C}}{\Msol} \right)^{0.60}
\end{equation}
and
\begin{equation}
\rm{\hat{SFE}} \simeq 0.96 \left( \frac{\hat{\rm{M}}_{C}}{\Msol} \right)^{-0.40}
\end{equation}
reproduce the main trends of the plots precisely with parameter fit errors well below $1\%$. In all simulations, the SFR increases towards the end of the cloud lifetime resulting in typical age distribution diagrams such as fig.~\ref{fig:parstud_age_diag}: most stars form just before star formation is terminated. Compared to the \citet{Zamora2012} model, we find a similar peak SFR for the $R_{\rm 0} = 64~\pc$ cloud of $\sim 500~\Msol / \yr$ (see their fig. 4).
\bef
\centering
\resizebox{\hsize}{!}{\includegraphics{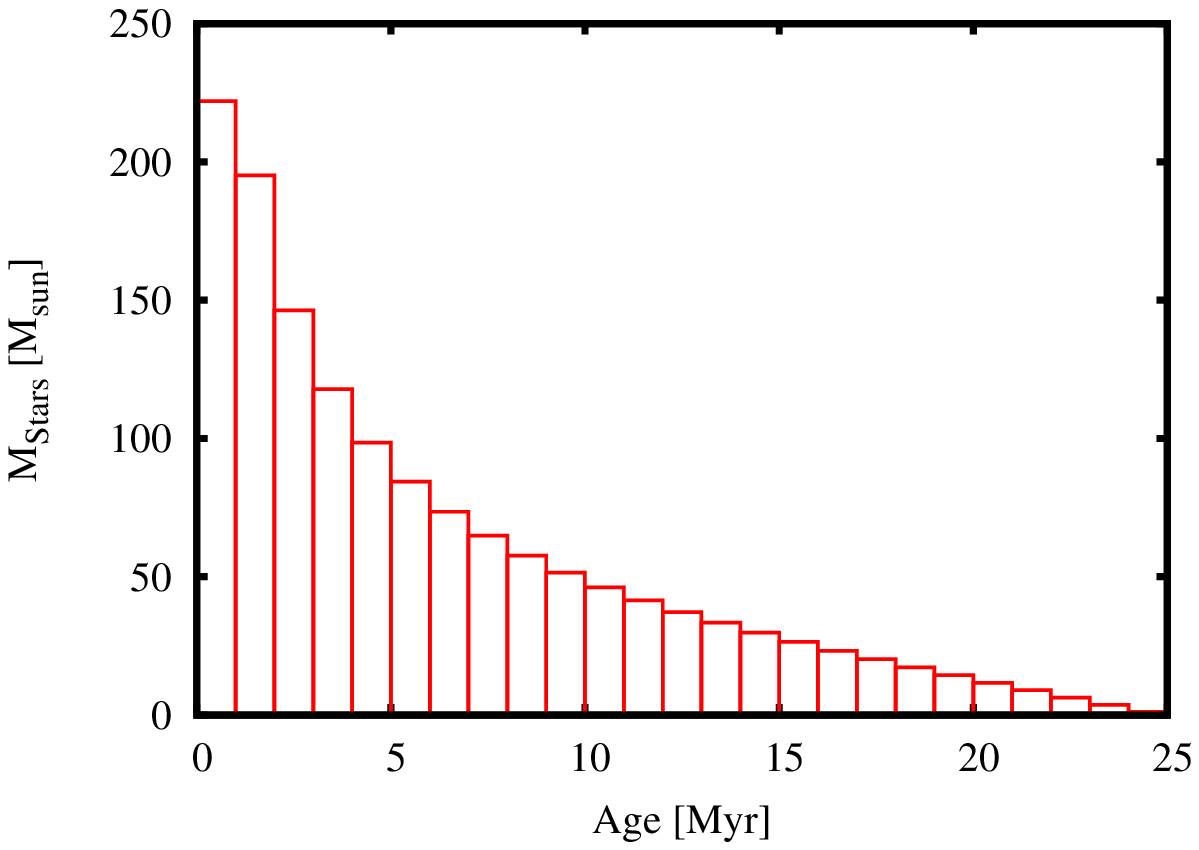}}
\caption{A typical distribution of stellar ages. Here, we selected $R_{\rm 0} = 64~\pc$ as a working example. The SFR peaks at the end of the cloud lifetimes for all clouds in this parameter study. Consequently, most stars are of young age.}
\label{fig:parstud_age_diag}
\eef
For the total number of massive stars, we find a power law
\begin{equation}
N_{\rm OB} \simeq 0.016 \left( \frac{\hat{\rm{M}}_{\rm C}}{\Msol} \right)^{0.58} 
\end{equation}
with the smallest clouds forming just one massive star, while clouds with peak masses of $10^{6}~\Msol$ form more than 50 massive stars. \citet{Zamora2012} found a number of $\sim 100$ massive stars for a $R_{\rm 0} = 100~\pc$ cloud. For the same cloud, our model predicts the total number of massive stars that formed over the whole cloud lifetime to be $\sim 30$. 
\subsection{Supernova activity}
In fig.~\ref{fig:parstud_snae} we present an overview of the cloud lifetime and supernova activity as predicted by the model.\\
Our model predicts that intermediate-size clouds in the range of $10^{3}~\Msol$ to $10^{4}~\Msol$ feature the shortest life expectancy with a minimum just below $20~\Myr$, while both larger and smaller clouds live longer lifes peaking at $30~\Myr$.
Regarding supernovae, the smallest clouds considered in this parameter study form just enough stars to produce slightly less than one massive star that leaves the main sequence and goes off as supernova. Strictly speaking, a result of less than one SN must be interpreted as some probability to form a massive star - or not. More massive clouds, on the other hand, form enough massive stars to see one and more supernovae in the second half of their lifetimes. The most massive clouds of the sample ($>10^{5}~\Msol$) see more than 30 supernova explosions, even more massive clouds are likely to see more as they will have greater lifetimes and higher stellar content (cf. fig.\ref{fig:parstud_snae}).\\
%
%
\bef
\centering
\resizebox{\hsize}{!}{\includegraphics{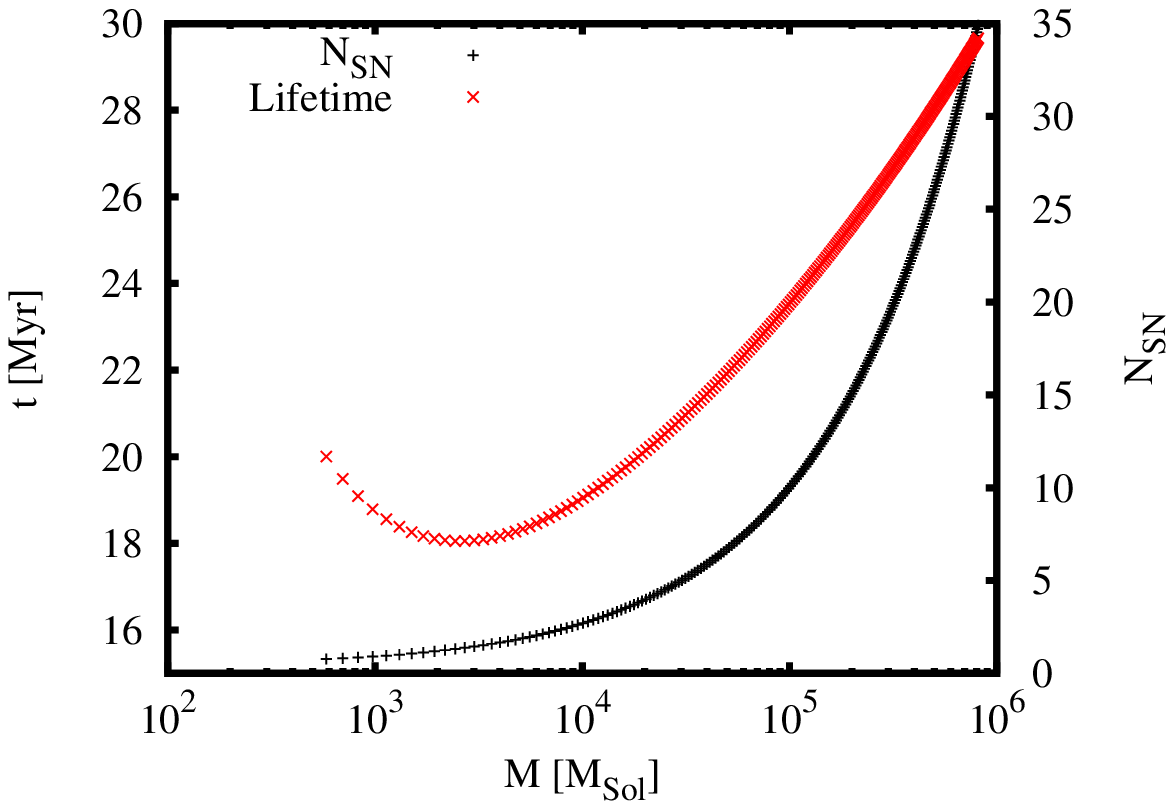}}
\caption{Supernova activity for clouds with varying peak masses. In the model, clouds below a peak mass of $\sim 10^{3}~\Msol$ do not form enough stars and have too small lifetimes to create at least one massive post main sequence star that might blast off as a supernova. Beyond that, more massive clouds are subject to an increasing number of supernovae.}
\label{fig:parstud_snae}
\eef
\section{Conclusions and discussion}
%
Step by step, we derived the key ingredients to construct a full (semi-)analytical model to describe the dynamical evolution of molecular clouds, the principle stellar nurseries of our universe. At each step, we tried to stick to observations and fundamental astrophysics as close as possible and did not introduce any additional efficiencies, parameters or other \textit{fudge factors}.\\
We started with the global evolution of the cloud, adopting the picture of clouds that can stay in a virialized equilibrium until they start to collapse. The clouds are fed via accretion from two gas streams, but the model does not depend on such an interpretation. Molecular gas forms and is transformed via a star formation criterion that incorporates the idea that the cloud maintains a distribution of cores that follows a Salpeter law. From that, the star forming mass fraction can be determined. Finally, the stars are allowed to feedback with their environment via ionization and supernovae enabling us to calculate evolutionary paths of molecular clouds for a wide variety of cloud sizes.\\
The main results may be summarized and compared with other work and observational results as follows.
\subsection{Cloud evolution}
Molecular clouds evolve over time. Even clouds similar in mass or radius may exhibit different levels of star formation activity simply because they are in different evolutionary phases. This resembles observational results by \citet{Kawamura2009} who defined different stages of molecular cloud evolution.\\
The typical lifetimes of clouds of sizes between a few parsecs to a few hundred parsecs are $20-30~\Myr$ resembling results by \citet{Murray2011} who found very specific lifetimes of $27\pm13~\Myr$. Continuous accretion and star formation delay the collapse and extend their lifetimes beyond the free-fall time. In addition, small clouds with $R_0 < 10~\pc$ can accrete gas for up to $20~\Myr$ before global collapse sets in, representing the fact that smaller clouds are more likely to be found in virial equilibrium \citep[see,e.g.][]{Heyer2001}.
\subsection{Star formation}
The CMF star formation criterion yields results comparable to established star formation criteria such as \citet{Krumholz2005} and shows similar scaling with cloud mass and turbulent Mach number. In contrast to the Krumholz model that relies on a number of efficiencies determined by numerical experiments, the CMF criterion incorporates only one observational parameter which is the Salpeter slope.
Further, the CMF criterion predicts non-universal star formation threshold densities of the order of $10^{4}~\pcc$. Observational studies by various authors \citep[see,e.g.][]{Lada2010, Pudritz2013, Rathborne2014} confirm that. Regarding turbulence, the CMF criterion is not very sensitive to increasing cloud Mach numbers, showing just a weakly negative scaling. However, it predicts that beyond a certain Mach number star formation is terminated.
The idea that the star formation rate is simply given as dense gas content over cloud free-fall time \citep[see,e.g.][]{Krumholz2005} results in SFRs in the range $10-1400~\Msol/\Myr$ for clouds in the range $10^{3}-10^{6}~\Msol$. A fiducial power law for the gas accretion velocity used in the paper is presented in sec.~\ref{sec:parstud_sfre}, different values result in slightly different slopes. The predicted SFRs compare well with observational results by \citet{Murray2011} or \citet{Lada2010}. For example, they found $715~\Msol/\Myr$ for \textit{Orion A}, a well evolved GMC with a mass of roughly $10^{5}~\Msol$ \citep{Lada2010}. Younger clouds of comparable mass such as \textit{California} or \textit{Orion B} are characterized by a much smaller stellar content, implying they are in an earlier phase of their evolution. Hence, their star formation rates are much smaller. In principle, the model presented here can be applied to such objects in order to derive an age estimate.
Further, all simulations of individual clouds have a star formation rate that increases over time and leads to the typical picture of stellar clusters that are dominated by young stars. That fits the results given by \citet{Palla2000} who found that most molecular clouds must have undergone a phase of accelerating star formation to reproduce the stellar cluster age histograms we observe today.\\
%
%
%
%
%
%
For the smallest clouds, we find star formation efficiencies of up to $10~\%$ while the most massive clouds peak at SFEs slightly below $1~\%$. Both the numerical values and the scaling with cloud mass fit observational findings \citep[see,e.g.][]{Krumholz2005, Murray2011}. However, the $\rm{SFE}$ as employed in this paper also accounts for molecular material that has been destroyed via feedback processes. The naive approach of an star formation efficiency defined as $M_{\rm S} / M_{\rm C}$ leads to doubled values, resembling the observations even better.\\
\subsection{Feedback}
Within their range of validity, both feedback effects have just small impact on the evolution of the cloud and decrease the peak star formation rates and efficiencies just by a few percent. Partly, this behavior can be attributed to an inherent limitation of the feedback effects to molecular gas destruction within some given volume fraction of the cloud. Further, we assume those cloud fractions to be filled with mean density molecular gas which does not account for the filamentary structure of molecular clouds and the high density of the star forming cloud parts. As a result, our feedback description may underestimate the overall feedback efficiency and a better description needs to account for the density structure of the cloud as well as the ionized regions \citep[see,e.g.][]{Franco2000, Peters2010b}. The energy and momentum input via outflows, stellar winds and supernovae was neglected in this model, but it may have an effect on the cloud's Mach number and modulate the star formation history once a sufficient number of stars emerged \citep[see,e.g.][]{Rogers2013, Bally2016, Padoan2016, Wareing2017}. 
On the other hand, our model employs the assumption that all HII regions within the cloud are bound to massive stars which may not hold true depending on the cloud structure and the location of the ionizing star which may lead to higher fractions of ionized gas even up to total cloud ionization \citep[see,e.g.][]{Franco1990}. For small clouds, our scheme may potentially underestimate the impact of ionization. Apart from photoionization, molecular gas destruction via photodissociation driven by lower-mass stars may serve as an additional feedback effect \citep[see,e.g.][]{Diaz1998}.\\
To conclude, In this model star formation is regulated neither by turbulence, nor by any stellar feedback effect. Rather, the star formation criterion itself limits the mass that is converted into stars, feedback processes are only important to limit the lifetime of the cloud. The star formation criterion itself is self-regulating and does not tend to convert clouds entirely into stars even if implemented without competing feedback effects. This result is consistent with the picture of \textit{fragmentation induced starvation}, i.e. a limited star formation efficiency due to fragmentation into subcritical objects \citep{Peters2010, Girichidis2012}.\\
\subsection{Conclusions}
The (semi-)analytical model presented in this paper agrees well with a large number of observational findings. Molecular cloud evolution can be described by an interplay between accretion, global contraction, star formation and (modest) feedback effects which can be modelled and understood using basic astrophysics and established astronomical constraints.\\
\begin{acknowledgements}
We would like to thank the anonymous referee for helpful comments that improved our manuscript. RB and BK acknowledge funding for this project by the DFG via the ISM-SPP 1573 grants BA 3706/3-1 and BA 3706/3-2, as well as for the grants BA 3706/4-1 and BA 3706/15-1. 
\end{acknowledgements}

\bibliographystyle{aa}
\bibliography{astro.bib}

\end{document}